\documentclass[aps,prl,reprint,showpacs,superscriptaddress]{revtex4-1}

\usepackage{graphicx}

\begin{document}

\title{Mn$_2$FeSbO$_6$: a ferrimagnetic ilmenite and an antiferromagnetic perovskite}

\author{R. Mathieu}\email{roland.mathieu@angstrom.uu.se}
\affiliation{Department of Engineering Sciences, Uppsala University, Box 534, SE-751 21 Uppsala, Sweden}

\author{S. A. Ivanov}
\affiliation{Department of Engineering Sciences, Uppsala University, Box 534, SE-751 21 Uppsala, Sweden}
\affiliation{Department of Inorganic Materials, Karpov' Institute of Physical Chemistry, Vorontsovo pole, 10 105064, Moscow K-64, Russia}

\author{I. V. Solovyev}	
\affiliation{National Institute for Materials Science, 1-2-1 Sengen, Tsukuba, Ibaraki 305-0047, Japan}

\author{G. V. Bazuev}
\affiliation{Institute of Solid-State Chemistry, Ural Branch of the Russian Academy of Science, 620999 Ekaterinburg, GSP-145, Russia}

\author{P. Anil Kumar}
\affiliation{Department of Engineering Sciences, Uppsala University, Box 534, SE-751 21 Uppsala, Sweden}

\author{P. Lazor}
\affiliation{Department of Earth Sciences, Uppsala University, Villav\"agen 16, SE-752 36 Uppsala, Sweden}

\author{P. Nordblad}
\affiliation{Department of Engineering Sciences, Uppsala University, Box 534, SE-751 21 Uppsala, Sweden}

\date{\today}

\begin{abstract}

Due of its polymorphism, Mn$_2$FeSbO$_6$ can be synthesized at high pressures and temperatures as a ferrimagnetic ilmenite or an antiferromagnetic perovskite. The structural phase transformation is discussed in detail, and magnetic structures are proposed for both phases. The high-pressure Mn$_2$FeSbO$_6$ polymorph is a rare example of $A_2 B'B''$O$_6$ perovskite with solely Mn cations on the $A$-site. Fe and Sb cations are ordered on the $B$-sites. Theoretical calculations for  the perovskite phase suggest a complex magnetic structure, holding an electronic polarization.

\end{abstract}


\maketitle

There is an ongoing quest for new materials showing several, possibly interacting, ferroic properties.  Such multiferroic materials may e.g. exhibit ferromagnetism and ferroelectricity\cite{reviews}, as well as magnetoelectric effects related to the correlation of spin and dipole ordering\cite{fiebig,tokura,dd}. Multiferroic and/or magnetoelectric materials have fundamental and applied importance in the fields of strongly correlated transition metal oxides and spintronics. A strategy to obtain such new materials is to design complex crystal structures including several magnetic cations, and small diamagnetic cations favoring polar distortions such as Te$^{6+}$, Nb$^{5+}$ or Sb$^{5+}$ [\onlinecite{CTO}]. 

Mn$_2$FeSbO$_6$ (MFSO) ilmenite-type mineral, called melanostibite\cite{mfso-apl}, contains two magnetic ions (Mn$^{2+}$ and Fe$^{3+}$) on different crystallographic sites. MFSO cannot be synthesized under ambient pressure. Yet it was shown that MFSO ilmenite could be synthesized with relatively high purity using thermobaric treatments\cite{mfso-apl,Bazuev}. MFSO ilmenite was found to order ferrimagnetically below 270 K.  By increasing the pressure employed during the thermobaric synthesis, a double perovskite phase of MFSO may be stabilized, with Mn cations on the $A$-site, and Fe, Sb cations ordered on the $B$-site. 

We show in the present article that the perovskite phase has magnetic properties quite different from those of the ilmenite one, namely is antiferromagnetic below 60 K. We discuss the phase transformation from ilmenite to perovskite in MFSO based on X-ray powder diffraction results. Using theoretical calculations, we predict a complex magnetic structure, and electronic polarization of the perovskite phase.\\

\begin{figure}[h]
\includegraphics[width=0.46\textwidth]{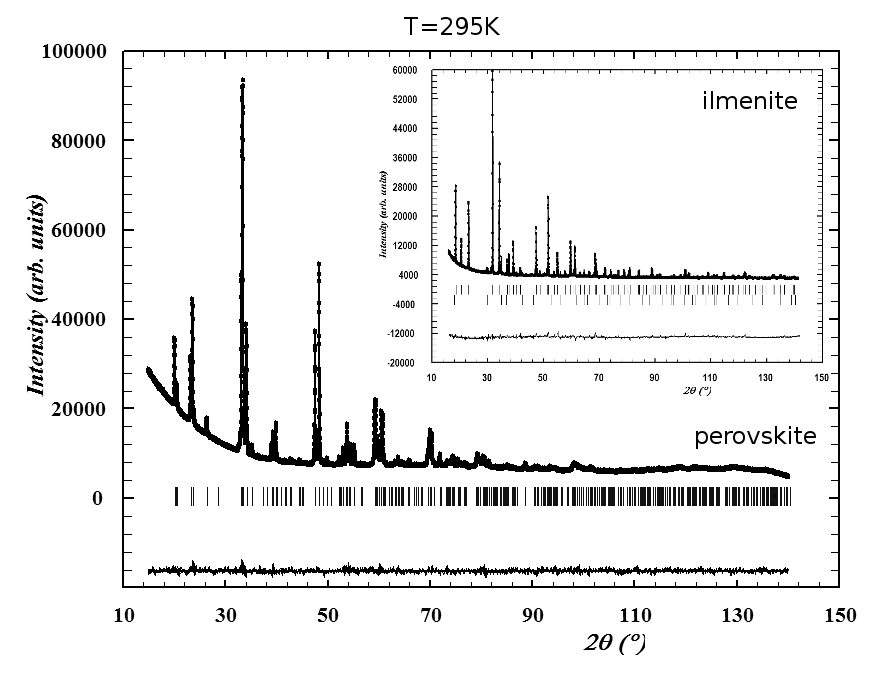}
\caption{X-ray diffraction patterns at 295 K and Rietveld refinements of perovskite (main frame; $R_p$ = 0.031, $R_{wp}$ = 0.049, and $R_b$ = 0.035) and ilmenite (inset; $R_p$ = 0.046, $R_{wp}$ = 0.067, and $R_b$ = 0.041) phases.} 
\label{fig-xrd}
\end{figure}

\begin{figure*}[t]
\includegraphics[width=0.9\textwidth]{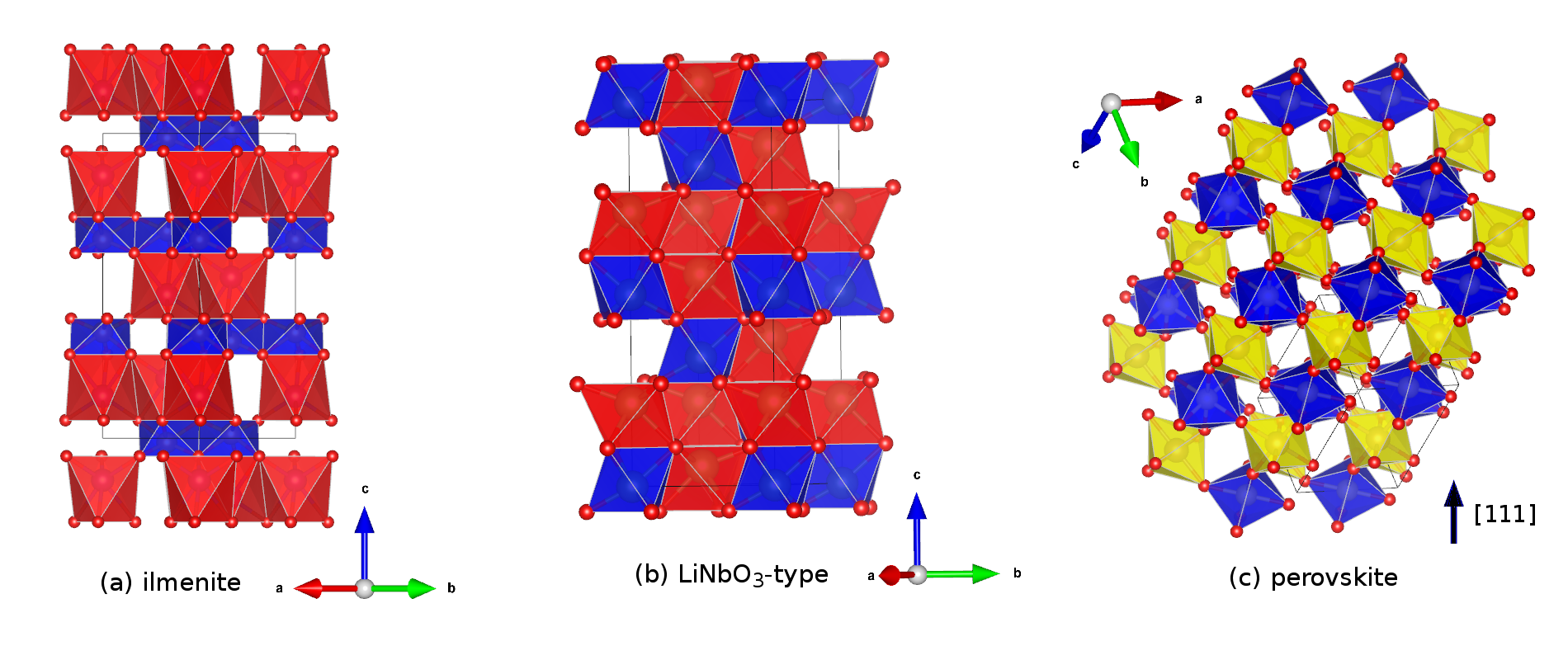}
\caption{(Color online) Polyhedral representations of different polymorphs in comparable orientation\cite{Megaw}. The color of the polyhedra reflects the central cation; (a,b) red: Mn, blue: Fe/Sb, (c) blue: Fe, yellow: Sb; Mn polyhedra are omitted for clarity. Small red spheres indicate oxygen atoms.}
\label{fig-struct}
\end{figure*}

\begin{table*}[t]
\caption{Structural parameters at room temperature. $V/Z$ is the unit cell volume per formula unit, $\rho$ is the density. Minimum-maximum values of $M$-O  bond lengths, $M$-O-$M$ bond angles, and $M$-$M$ distances ($M$ = Fe or Mn) are indicated.}
\label{table-param}

\begin{tabular}{c|c|c}
 & ilmenite & perovskite\\
\colrule
Space group& $R\bar{3}$&$P2_1/n$\\
a, b, c ({\AA}) ; $\beta$ (deg.) &  5.2379(4), 14.3519(6) &  5.2459(4), 5.3987(4), 7.6332(5) ; 89.67(6)\\
Coordination number  for Mn / Fe & 6 / 6 & 8(+4) / 6 \\
V/z ({\AA}$^3$) ; $\rho$ (g/cm$^3$)  &  56.8 ; 5.23 &  54.1 ; 5.89\\
Mn-Mn / Mn-Fe / Fe-Fe distances  ({\AA}) &   3.129-3.981 /  2.985-3.989 / 3.048 & 3.847-3.683 / 3.099-3.519 / 5.246-5.398\\
Mn-O / Fe-O distances ({\AA}) & 2.295-2.564 / 1.738-1.962  & 2.124-2.764 / 1.949-2.094 \\
Mn-O-Mn / Mn-O-Fe / Fe-O-Fe angles (deg.) &  60.6-151.6 / 58.8-145.3 / 55.3-136.1 & 78.9-156.4 / 51.7-140.7 / 125.5-135.3\\
Polyhedral volume ({\AA}$^3$) ; distortion Mn/Fe-O$_6$ & 17.80 ; 0.043 / 7.72 ; 0.076  & 22.67 ; 0.067 / 10.87 ; 0.005\\
Mn / Fe cation shift from centroid ({\AA}) &  0.191 / 0.053& 0.161 / 0

\end{tabular}
\label{comp}
\end{table*}

Mn$_2$FeSbO$_6$ ilmenite, i.e. with the same structure as the mineral material, was synthesized as ceramic using the following thermobaric treatment: pressure $P$ = 3 GPa, temperature $T$ = 1000 $^o$C, duration $d$ = 30 mins (see Ref. \cite{mfso-apl,Bazuev} for details). By increasing the pressure to 6 GPa (and duration to 40 mins), a perovskite phase of MFSO could be stabilized \cite{Bazuev}. Crystal structure and stoichiometry of ilmenite and perovskite MFSO were investigated by X-ray powder diffraction (XRPD) on a D8 Bruker diffractometer using CuK$_{\alpha 1}$ radiation and microprobe energy-dispersive spectroscopy analysis. Magnetization and heat capacity measurements were performed using a superconducting quantum interference device (SQUID) MPMS magnetometer and PPMS physical property measurement system from Quantum Design Inc. Complementary high-temperature magnetization measurements were performed using the VSM/oven option of the PPMS. Electronic structure calculations were performed to predict the magnetic ordering of the perovskite phase. Crystal and some of the magnetic structures were drawn using VESTA\cite{vesta}.\\

The XRPD patterns shown in Fig.~\ref{fig-xrd} were obtained for ilmenite and perovskite compounds. The data sets were successfully refined using the Rietveld method, permitting the determination of the structural parameters listed in Table~\ref{table-param}. As shown in the table, the perovskite MFSO adopts a monoclinically distorted phase, with $\beta$ angle close to 90$^o$ (i.e. almost orthorhombic). The XRPD data evidences the ordering of the Fe/Sb cations on the perovskite $B$-site.

MFSO is dimorphic; the ilmenite phase is the low-pressure polymorph, while the perovskite one is the high-pressure one. The transformation of ilmenite into perovskite phase at high hydrostatic pressures is accompanied by a decrease in the cell volume and distances between the layers of the hexagonal anion packing, i.e. by the formation of a denser structure (see Table~\ref{table-param}). From a crystallochemical point of view, the reconstructive phase transition from ilmenite to perovskite phase in MFSO occurs by incorporation of Mn cations from interlayer positions into the layers of the closest packing. This redistribution may occur since Mn$^{2+}$ cations have no site preference, and can occupy sites of various coordination. The hexagonal type of packing then changes to the cubic one.

The ilmenite structure is based on a hexagonal close-packed array of oxygen. Cations occupy 2/3 of the octahedral sites leaving 1/3 of the octahedral sites vacant. Mn and Fe cations are distributed between alternating layers such that each Mn octahedra shares a face with the Fe octahedra above or below (but not both) and also shares edges with three other Mn octahedra in the same layer (see Fig.~\ref{fig-struct}). The Fe octahedra have similar linkages. The octahedral holes of the oxygen array are linked by face-sharing into chains along the $c$-axis that are filled in the sequence vacancy-Mn-Fe-vacancy-Fe-Mn-vacancy, etc. In the (0001) planes, the ilmenite structure has the same type of cations. We can assume that in the first stage of the high-temperature/high-pressure-induced phase transformation, the ilmenite structure is turned into the LiNbO$_3$-type structure (Mn-Fe-vacancy-Mn-Fe-vacancy-Mn-Fe-vacancy, etc) in which two different cations (Mn and Fe) are arranged alternatively in (0001) planes\cite{Megaw} (see Fig.~\ref{fig-struct}). This mechanism requires Mn and Fe cation rearrangement and is accompanied by a topological change in the structure to accommodate cation redistribution. Several possibilities may be proposed to describe those cation displacements: exchange of two edge-shared cations (using the vacant site) and two face-shared cations, diffusion of cations and octahedral tilting. The cation diffusion involves breaking bonds, which requires high temperatures due to high activation energy. The octahedral tilting may help lowering the activation energy creating of the necessary topological accommodation for Mn and Fe cations during the phase transition. The perovskite structure is derived from the ideal cubic $Pm-3m$ structure by a combination of in-phase (resp. out-of-phase) octahedral tilting about $<$001$>$ (resp. $<$110$>$) axis of the prototype perovskite structure. The larger Mn cations occupy the cuboctahedral sites, although as a consequence of the tilting of the FeO$_6$ octahedra (around 19 deg.). The Mn cations are effectively eight coordinated with 4 more oxygen anions located at greater distances (more than 3 {\AA}). The Mn-O bond length increases in the transformation from ilmenite to perovskite, increasing the Mn coordination. The Fe coordination however does not change.\\

\begin{figure}[h]
\includegraphics[width=0.46\textwidth]{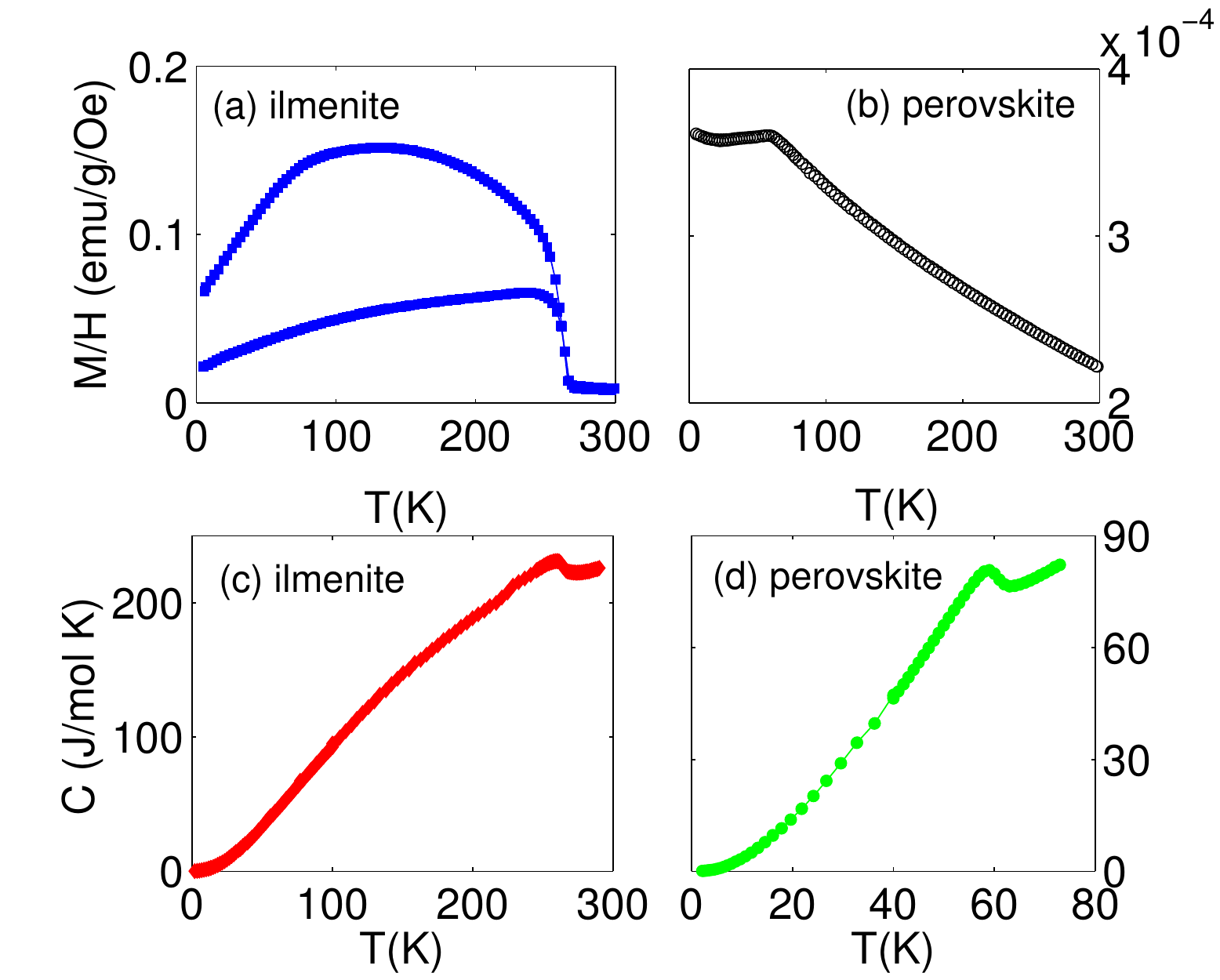}
\caption{(Color online) Upper panels: Zero-field-cooled (ZFC) and field-cooled (FC) magnetization curves for MFSO in (a, $H$ = 20 Oe) ilmenite and (b, $H$ = 3 T) perovskite phases. The magnetization of  MFSO perovskite was recorded in a larger field as the  low-field magnetic response of the perovskite phase is more affected by minor amounts of magnetic impurities. In that field, ZFC and FC curves closely coincide with each other. Lower panels: Temperature dependence of the heat capacity $C$ of MFSO in (c) ilmenite and (d) perovskite phases.}
\label{fig-MTHC}
\end{figure}

Figure~\ref{fig-MTHC} (upper panels) shows the temperature $T$ dependence of the magnetization $M$ (plotted as $M/H$) for both MFSO compounds. While the ilmenite compound displays a ferrimagnetic transition near 270 K\cite{mfso-apl}, perovskite MFSO (lower panel) exhibits a paramagnetic response near that temperature.  On the other hand a peak is observed in the magnetization curves near 60 K, suggesting antiferromagnetic ordering below that temperature. Magnetic hysteresis measurements performed at low temperatures ($T$ = 10 K, not shown) indicate a linear response to magnetic field in perovskite MFSO, in contrast to the ilmenite case.\cite{mfso-apl}
 
\begin{figure*} [t]
\includegraphics[width=0.15\textwidth]{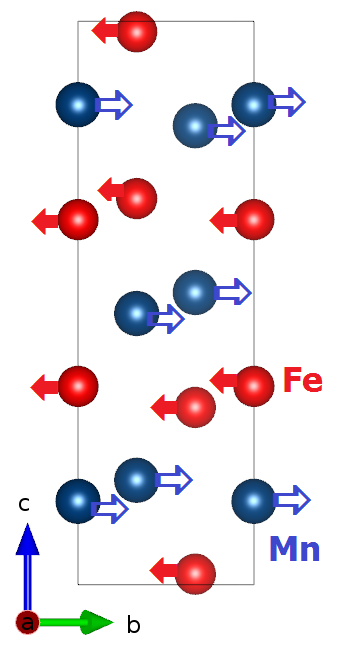}
\includegraphics[width=0.43\textwidth]{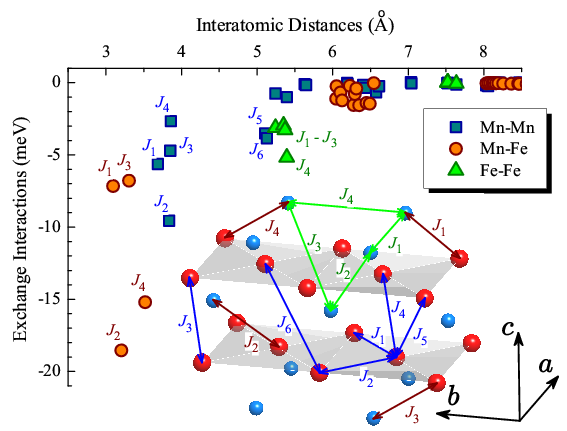}
\includegraphics[width=0.35\textwidth]{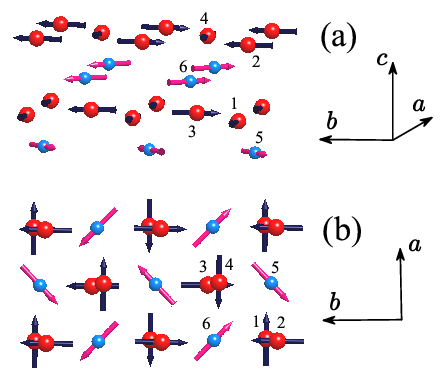}
\caption{(Color online) Left panel: Predicted magnetic structure for (left) ilmenite and (right) perovskite MFSO; (a) side view and (b) top view. Mn and Fe atoms are indicated in red and blue colors respectively. The atoms, forming the primitive cell of Mn$_2$FeSbO$_6$, are indicated by numbers. The middle panel shows the distance dependence of interatomic exchange interactions within Mn and Fe sublattices (squares and triangles, respectively), and between Mn and Fe sublattices (circles); see inset for notation. The interactions between other sites can be obtained by applying the symmetry operations of the space group $P2_1/n$.}
\label{fig-magstruct}
\end{figure*}

The inverse susceptibility of the perovskite material does not exactly follow a Curie-Weiss behavior above 100 K; however considering a near-linear behavior of $H/M$ data measured at higher temperatures (300 K $<$ $T$ $<$ 500K, not shown) yields a Curie-Weiss temperature $\theta_{CW}$ $\sim$ - 180 K. The respective magnetic transitions of the two MFSO materials are also observed in the heat capacity $C$ curves depicted in Fig.~\ref{fig-MTHC}, at 268 K and 59 K respectively.\\

One oxide with ilmenite structure is the quasi two-dimensional antiferromagnet MnTiO$_3$\cite{MTO}. In that structure, the magnetic moments of Mn$^{2+}$ cations are antiferromagnetically coupled to each other within and in between the hexagonal (0001) planes. Interestingly, replacing Mn$^{2+}$ by Fe$^{2+}$ as in FeTiO$_3$ brings forth a ferromagnetic coupling within the hexagonal planes, although adjacent planes remain antiferromagnetically coupled\cite{FTO}. The NiMnO$_3$ and CoMnO$_3$ ilmenites\cite{FMI} instead display ferromagnetic (ferrimagnetic) ordering above room temperature. The microscopic magnetic structure of NiMnO$_3$ was investigated in detail using neutron diffraction\cite{NMO}. It was found that the ferrimagnetic state is composed of alternating Ni$^{2+}$ and Mn$^{4+}$ hexagonal planes, yielding an interaction pattern between magnetic cations similar to that of hematite Fe$_2$O$_3$. We could not perform neutron diffraction experiments on our samples since less than 100 mg of each material were synthesized, and  Yet, if we similarly transpose the crystal and magnetic structure of NiMnO$_3$ to the one of ilmenite MFSO, replacing in their respective atomic positions Mn$^{4+}$ and Ni$^{2+}$ in NiMnO$_3$ by Mn$^{2+}$ and Fe$^{3+}$/Sb$^{5+}$ in MFSO, we can propose the magnetic structure presented in Fig.~\ref{fig-magstruct}. This structure is also reminiscent of that of the Fe$_2$O$_3$ hematite, as predicted theoretically\cite{mfso-apl}.\\

Let us now consider the perovskite phase of Mn$_2$FeSbO$_6$; there are rather few perovskites with solely Mn cations on the perovskite A-site. Relatively large amounts of Mn could be doped onto the $A$-site under pressure:  33 \% in (In,Mn)MnO$_3$ ($P$ = 6 GPa, $T$ = 1500 $^o$C, $d$ = 40 mins), and 75 \% in LaMn$_3$Cr$_4$O$_{12}$ ($P$ = 8-10 GPa, $T$ = 1000 $^o$C, $d$ = 30 mins), bringing forth antiferromagnetic monoclinic\cite{IMMO} resp. cubic\cite{LMCO} phases. It seems that only one perovskite compound with solely A-site Mn cations has been reported so far: MnVO$_3$ synthesized under pressure\cite{MVO}. Akin to MFSO, the structure of MnVO$_3$ (MVO) changes from ilmenite to perovskite above a critical pressure of  $\sim$ 4.5 GPa. In the high-pressure conditions ($P$ $>$ 4.5 GPa, $T$ $\sim$ 1000-1200$^o$C, $d$ $\sim$ 30-60 mins), an orthorhombic MVO phase is stabilized. In constrast to MFSO, the magnetic properties of ilmenite and perovskite phases of MVO were found to be quite similar, with a weak ferromagnetism below 65-70 K reported in both cases\cite{MVO}.  $\theta_{CW}$ was found to decrease (in absolute value) from $\sim$ -430 to -250 K between ilmenite and perovskite phases\cite{MVO}. A recent study instead reported for perovskite MVO synthesized under pressure ($P$ = 8 GPa, $T$ = 1100 $^o$C, $d$ = 30 mins), antiferromagnetism below 46K, and $\theta_{CW}$ $\sim$ -154K\cite{MVO-attfield}.\\

We investigate in more detail the magnetic configuration of the MFSO compounds in their respective phases. As discussed above, by analogy with the NiMnO$_3$ ilmenite, we can predict the magnetic structure of MFSO ilmenite (see left panel of Fig.~\ref{fig-magstruct}). The structure matches the results of  earlier theoretical calculations, in which Fe and Mn cation were ferromagnetically coupled within their respective hexagonal planes, and Fe and Mn cations were antiferromagnetically coupled along the $c$-direction\cite{mfso-apl}. In the perovskite case, we have elaborated another theoretical model to find the possible magnetic structures.  For these purposes, we construct an electronic low-energy (Hubbard-type) model for the `magnetic' $3d$ bands of Mn and Fe, using results of first-principles electronic structure calculations and the experimental parameters of the crystal structure. The advantage of the model analysis is that it is relatively easy to consider many types of different magnetic structures and, thus, determine the correct magnetic ground state, along the same line as it was recently done for multiferroic manganites~\cite{IgorMF}. More specifically, we start with the local-density approximation (LDA), and determine parameters such as crystal field, transfer integrals and effective Coulomb interaction in the $3d$ bands. Since the $3d$ bands are well separated from the rest of the spectrum, such a construction is rather straightforward (the details can be found in Ref.~\cite{review2008}, and the parameters are listed in Ref~\cite{SM}). Then, we solve the obtained model in the mean-field Hartree-Fock (HF) approximation and derive parameters of interatomic exchange interactions, using the method of infinitesimal spin rotations near an equilibrium magnetic state~\cite{review2008}. The behavior of these parameters is depicted in the middle panel of Fig.~\ref{fig-magstruct}, and the magnetic model itself is defined as ${\cal H}_S = -\sum_{\langle ij \rangle} J_{ij} {\bf e}_i \cdot {\bf e}_j$, where ${\bf e}_i$ is the \textit{direction} of the spin at the site $i$, and the summation runs over the \textit{inequivalent pairs} $\langle ij \rangle$. As expected for the $d^5$ configuration of the Mn$^{3+}$ and Fe$^{2+}$ ions, all interactions are antiferromagnetic (AFM). Moreover, the structure of these interactions is very complex: the interactions spread far beyond the nearest neighbors and compete with each other. This partly explains the relatively large ratio of $|\theta_{CW}|$ to the magnetic transition temperature, and in comparison to the ilmenite phase.

In order to determine the magnetic structure, resulting from competing AFM interactions, we have performed extensive HF calculations of the low-energy model. First, we searched for the lowest energy structure with collinear arrangement of spins within the primitive cell. We have found that this magnetic structure is $\uparrow \uparrow \downarrow \downarrow \downarrow \uparrow$, where the arrows describe the relative directions of spins at the site 1-6 (see right panel of Fig.~\ref{fig-magstruct} for the notations). Then, we considered possible modulations of this structure with the spin-spiral vector $q$. We have found that even without relativistic spin-orbit (SO) interaction, the $\uparrow \uparrow \downarrow \downarrow \downarrow \uparrow$ structure is unstable with respect to the spin spiral alignment and the theoretical magnetic ground state is close to $q$ = (0,1/2,0), in units of reciprocal lattice translations (the energy gain is about 23 meV per Mn$_2$FeSbO$_6$). The SO interaction specifies the spacial orientation of spins and align them mainly in the $ab$ plane, as shown in Fig.~\ref{fig-magstruct}. 

Although the magnetic spin-spiral alignment formally breaks the inversion operation, without SO interaction, the latter can always be combined with an appropriate spin rotation. Therefore, the electronic polarization ($P$) will identically vanish. In order to fully break the inversion symmetry, one should forbid the spin rotations. This is done by including the SO interaction, which thus yields finite $P$. Using the Berry-phase formalism, adopted for the low-energy model~\cite{IgorMF}, the latter can be estimated as $P \sim 10 \mu$C/m$^2$ and is mainly parallel to the $a$ axis. \\

In conclusion, we have investigated the magnetic properties of two polymorphs of Mn$_2$FeSbO$_6$,  namely the ferrimagnetic ilmenite one and the high-pressure antiferromagnetic perovskite. The latter phase is a rare example of $A_2B'B''$O$_3$ perovskite material with Mn cations on the $A$-site of the perovskite structure and, since Fe and Sb cations order on the $B$-sublattice, magnetic cations are present on both $A$- and $B$-sites. Theoretical calculations predict a complex magnetic structure, and electronic polarization for the perovskite phase. Interestingly, it was shown recently that high-pressure LiNbO$_3$-type MnTiO$_3$ (theory\cite{MTO-P}) and FeTiO$_3$ (experiments\cite{FTO-P}) acquire a spontaneous electronic polarization, suggesting the potential multiferroic nature of this family of materials.\\

\begin{acknowledgments}

We thank the Swedish Research Council (VR), the G\"oran Gustafsson Foundation, the Swedish Foundation for International Cooperation in Research and Higher Education (STINT), and the Russian Foundation for Basic Research for financial support. The work of IVS is partly supported by the grant of the Ministry of Education and Science of Russia N 14.A18.21.0889.

\end{acknowledgments}


\begin{thebibliography}{1}

\bibitem{reviews}
D. I. Khomskii, J. Magn. Magn. Mater. \textbf{306}, 1 (2006); S. -W. Cheong and M. Mostovoy, Nature Mater. \textbf{6}, 13 (2007); J. F. Scott, Nature Mater. \textbf{6}, 256 (2007).

\bibitem{fiebig}
M. Fiebig, Th. Lottermoser, D. Fr\"ohlich, and R. V. Pisarev, Nature \textbf{419}, 818 (2002).

\bibitem{tokura}
Y. Tokura and S. Seki, Adv. Materials \textbf{22}, 1554 (2010)N. A. Spaldin and M. Fiebig, Science  \textbf{309}, 391 (2005); 

\bibitem{dd}
D. Choudhury, P. Mandal, R. Mathieu, A. Hazarika, S. Rajan, A. Sundaresan, U. W. Waghmare, R. Knut, O. Karis, P. Nordblad, and D. D. Sarma, Phys. Rev. Lett. \textbf{108}, 127201 (2012).

\bibitem{CTO}
S. A. Ivanov, R. Tellgren, C. Ritter, P. Nordblad, R. Mathieu, G. Andr\'e, N. V. Golubko, E. D. Politova, and M. Weil, Mater. Res. Bull. \textbf{47}, 63 (2012).

\bibitem{mfso-apl}
R. Mathieu, S. A. Ivanov, G. V. Bazuev, M. Hudl, P. Lazor, I. V. Solovyev, and P. Nordblad, Appl. Phys. Lett. \textbf{98}, 202505 (2011).

\bibitem{Bazuev}
G. V. Bazuev, B. G. Golovkin, N. V. Lukin, N. I. Kadyrova, and Yu. G. Zainulin. J. Solid State Chem. \textbf{124}, 333 (1996); A. P. Tyutyunnik, G. V. Bazuev, M. V. Kuznetsov, Yu.G. Zainulin, Mater. Res. Bull. \textbf{46}, 1247 (2011).

\bibitem{vesta}
K. Momma and F. Izumi, J. Appl. Crystallogr. \textbf{41}, 653 (2008).

\bibitem{Megaw}
H. Megaw, in: ``Crystal Structures; A Working Approach'' (W. B. Saunders, Philadelphia, 1973).

\bibitem{MTO}
 J. Akimitsu and Y. Ishikawa, J. Phys. Soc. Jpn. \textbf{42}, 462 (1977).

\bibitem{FTO}
H. Kato, Y. Yamaguchi, M. Ohashi, M. Yamada, H. Takei, S. Funahashi, Solid State Comm. \textbf{45}, 669 (1983).

\bibitem{FMI}
T. J. Swoboda, R. C. Toole, and J. D. Vaughan, J. Phys. Chem. Solids  \textbf{5}, 293 (1958). 

\bibitem{NMO}
M. Pernet, J. C. Joubert, and B. Ferrand, Solid State Comm. \textbf{16}, 503 (1975); (in French). 

\bibitem{IMMO}
A. A. Belik, Y. Matsushita, M. Tanaka, and E. Takayama-Muromachi, Angew. Chem. Int. Ed. \textbf{49}, 7723 (2010).

\bibitem{LMCO}
Y. Long, T. Saito, M. Mizumaki, A. Agui, and Y. Shimakawa, J. Am. Chem. Soc. \textbf{131}, 16244 (2009).

\bibitem{MVO}
Y. Syono, S.-I. Akimoto, and Y. Endoh, J. Phys. Chem. Solids \textbf{32}, 243 (1971).

\bibitem{MVO-attfield}
M. Markkula, A. M. Arevalo-Lopez, A. Kusmartseva, J. A. Rodgers, C. Ritter, H. Wu, and J. P. Attfield, Phys. Rev. B \textbf{84}, 094450 (2011).

\bibitem{IgorMF}
I.~V. Solovyev,
Phys. Rev. B \textbf{83}, 054404 (2011);
I.~V. Solovyev, M.~V. Valentyuk, and V.~V. Mazurenko, \textit{ibid.} \textbf{86}, 144406 (2012).

\bibitem{review2008}
I.~V. Solovyev,
J. Phys.: Condens. Matter \textbf{20}, 293201 (2008).

\bibitem{SM}
Supplemental materials [parameters of the crystal field, transfer integrals, and matrices of Coulomb interactions]. 

\bibitem{MTO-P}
X. Deng, W. Lu, H. Wang, H. Huang, and J. Dai, J. Mater. Res. \textbf{27}, 1421 (2012).

\bibitem{FTO-P}
T. Varga, A. Kumar, E. Vlahos, S. Denev, M. Park, S. Hong, T. Sanehira, Y. Wang, C. J. Fennie, S. K. Streiffer, X. Ke, P. Schiffer, V. Gopalan, and J. F. Mitchell, Phys. Rev. Lett. \textbf{103}, 047601 (2009).

\end{thebibliography}
\end{document}